\documentstyle[12pt]{article}
\title{Isotope effects in multiband superconductors}
\author{N.Kristoffel$\sp{a,b}$ and P.Rubin$\sp a$\\
$\sp a$Institute of Physics, University of Tartu,\\
 Riia 142, 51014 Tartu, Estonia\\
$\sp b$Institute of Theoretical Physics, University of Tartu,\\
T\"ahe 4, 51010 Tartu, Estonia}
\date{}
\begin{document}
\maketitle

\begin{abstract}
Isotope effects in a multiband superconductor with a leading interband pairing
channel are investigated. A relatively small electron-phonon contribution
into the pair-transfer interaction can cause effects of observed magnitude.
A multiband model which interpolates the cuprate properties is used for
illustrative calculations. Isotope exponents of the transition temperature
($\alpha$), supercarrier density ($\alpha_n$), paired carrier effective mass
($\alpha_m$) and of the penetration depth ($\alpha_{\lambda}$) on the
doping scale have been obtained. The known opposite trend of $\alpha$ and
$T_c$ is reflected. The trends of $\alpha$ and $\alpha_n$ are similar.
In contrast with the usual assumption in the interband case $\alpha_{\lambda}$
is driven by $\alpha_n$, which overwhelmes $\alpha_m$.
\end{abstract}


\section{Introduction}
The isotope effect on the superconductivity transition temperature has
indicated the participation of the lattice vibrations in the pairing
mechanism. In simple metals the superconductivity appears as a pure
phonon-mediated intraband effect due by attractive Fermi-level attached
effective electron coupling. For complex materials the deviations from
the canonical BCS value $\alpha =0.5$ become especially exposed.

For cuprate high-temperature superconductors the isotope effect at
$O\sp{16}\rightarrow O\sp{18}$ substitution has been proven in a number of
investigations, e.g. [1-7]. An overall trend of the isotope effect
exponent $\alpha$ to enchance with lowering transition temperature and
{\it vice versa} appeared to be characteristic for cuprates. This behaviour
of $\alpha$ on the doping phase diagram has stimulated the search for
nonphononic pairing mechanisms. It seems that phonon mechanisms cannot
compete with electronic (charge or spin) pairing channels for cuprates.
However the superconductivity mechanism of cuprates remains still elusive.
In the last time multiband approaches with interband interactions seem
to become popular [8,9].

Sometimes one asks simply whether the electron-phonon interaction plays an
essential role in cuprate superconductors or not [6]. Cuprates are known to
be systems with strong electronic correlations and lattice structural
effects. Doping induces here phase separation and formation of a
defect-polaronic subsystem bearing the doped holes [10,11]. Indications of
the influence of phonons on various properties correlated with the
superconductivity are known [12,13]. So, the mentioned question must be
about the relative contribution of the electron-phonon interaction
(intra- and interband) into the cuprate pairing mechanism.

The development of isotope effect investigations concentrates on the
CuO$_2$-plane magnetic penetration depth $\lambda$ as $\lambda\sp{-2}$
represents the superfield stiffness. It includes the supercarrier number
density $N_s$ and the in-plane paired carrier effective mass in the ratio
$N_sm_{ab}\sp{-1}$. In the conventional theory $N_s$ is usually interpreted
as the total carrier density. Correspondingly one interpretes the
$\alpha_{\lambda}$ as representing the paired carrier effective mass
isotope effect [6-7], because the number of the normal state carriers does
not change by isotopic substitution [3]. Such neglecting of the isotope effect
contribution from $N_s$ into $a_{\lambda}$ cannot be justified in general.
In nonconventional pairing mechanisms only a part of normal state carriers
will be paired even at $T=0$. The isotope change of $N_s$ cannot be
ignored and can contribute to the understanding of the pairing mechanism.

This is the case in multiband superconductivity with interband pairing
interaction. And more. A moderate electron-phonon contribution in the
(interband) pairing interaction as compared with a nonphononic one does
not mean necessarily a small outcome in the isotope effect. In the
framework of the two-band superconductivity [14,15] with the interband
pair-transfer channel [16] it has been shown that a small ($<10\%$)
contribution of the interband electron-phonon coupling into the whole
interband coupling can cause a remarkable isotope effect [16-18]. The
magnitude of $\alpha$ has been found to vary with $T_c$ just in the same
opposite manner as observed in cuprates. The promising outcome of two-band
models for the explanation of isotope effect peculiarities has not been
found much attention, however see [9,18-21].

The authors of the present contribution have developed a simple descriptive
(multiband) model of cuprate superconductivity [22,23]. It is based on
an electron spectrum created and evolving with doping. The interband
pair transfer channel operates between the itinerant and defect subsystem
states. This model explains qualitatively the behaviour of cuprate energetic
[23,24] and thermodynamic [25,26] characteristics on the whole doping scale.
In the present communication the isotope effect exponents are calculated using
this model. The characteristic behaviour of $\alpha$ with $T_c$ is obtained.
The supercarrier density isotope exponent follows the trend of $\alpha$.
The paired carrier effective mass isotope exponent is negative at underdoping.
Contrary to usual assumptions the superfluid density isotope effect is
determined by the supercarrier density change in collaboration with the
effective mass contribution.

\section{The isotope effect characteristics and the vibronic constant}
The isotope effect characteristic exponents are defined as
\begin{equation}
\alpha_{Xi}=-\frac{d\ln X}{d\ln M}C_i\; ,
\end{equation}
where $X$ designates the physical quantity under consideration and $i$
specifies the atomic mass ($M_i$) undertaken to be isotopically substituted.
The effective mass ($M$) attributed to the active vibration mode can be a
complicated function of the masses of the atoms in the unit cell. This is
accounted by $C_i=d\ln M(d\ln M_i)\sp{-1}$. The oxygen isotope effect in
cuprates can be considered as caused by the dominating $O\sp{16}\rightarrow
O\sp{18}$ mass change. In (1) for $T_c$ corresponds $\alpha$; $\alpha_n$ --
to supercarrier density $n_s$; $\alpha_m$ -- to supercarrier effective mass
$m_{ab}$, $\alpha_{\lambda}$ -- to the penetration depth $\lambda_{ab}$
and $\alpha_{\rho}$ -- to the superfluid stiffness $\rho_s \sim
\lambda_{ab}\sp{-2}$.

The penetration depth is expressed as
\begin{equation}
\lambda\sp 2=\frac{m_{ab}c\sp 2}{4\pi e\sp 2N_s}
\end{equation}
with $N_s=n_spN_0$, where $pN_0$ stands for the normal state carrier
concentration at hole doping $p$. One sees that
\begin{equation}
\frac{\alpha_m}{\alpha}=\frac{T_c}{m_{ab}} \frac{d m_{ab}}{d T_c}\; ,
\end{equation}
\begin{eqnarray}
\alpha_{\lambda} & = & \frac{1}{2}(\alpha_m-\alpha_n) \nonumber \\
\alpha_{\rho} & = & -2\alpha_{\lambda}\; .
\end{eqnarray}
The interelectronic effective coupling mediated by phonons of frequency
$\omega$ is characterized by
\begin{equation}
V_{\sigma\sigma '}(\vec{k},\vec{k}')=
\frac{|g_{\sigma\sigma '}(\vec{k}-\vec{k}')|\sp 2\hbar\omega (\vec{k}-\vec{k}')}
{[\epsilon_{\sigma}(\vec{k})-\epsilon_{\sigma '}(\vec{k}')]\sp 2 -
[\hbar\omega (\vec{k}-\vec{k}')]\sp 2}\; .
\end{equation}
Here $g_{\sigma\sigma '}$ are the coefficients of the linear
electron-phonon coupling.

Note that in the intraband case $\sigma =\sigma '$, as also in the
interband $\sigma \neq \sigma '$ case, the attractive part of the coupling
does not depend on $M$ because $g=\sqrt{\hbar/2M\omega_D}$ and $\omega_D
\sim M\sp{1/2}$ (in the BCS theory $M$ entres only through the integration
borders). In a large part of the $\vec{k}$ space, especially in the
interband case, where $\hbar\omega (\vec{k}-\vec{k}')$ can be neglected in
comparison with $|\epsilon_{\sigma}(\vec{k})-\epsilon_{\sigma '}(\vec{k})|$,
so that $V>0$ corresponds to repulsion, one has $V\sim M\sp{-1}$. In the
case of interband coupling the result $V_{\sigma ,\sigma '}\sim M\sp{-1}$
is of special significance because the repulsive interelectron coupling can
then lead to pairing [27] and $V$ can excert immediate influence on
superconductivity associated properties.

In the case of dominating interband pairing interaction the Expr.(1),
leaving out $C_i$, can be presented as
\begin{equation}
\alpha_X=Z\frac{W}{X}\frac{dX}{dW}\; ,
\end{equation}
where
\begin{equation}
Z=\frac{V}{W}
\end{equation}
determines the relative electron-phonon contribution to the whole interband
pairing interaction $W=V+U$ ($U$ is the Coulomb contribution).

\section{The model and the calculation scheme}
The physical model used [23] which interpolates the known cuprate properties
can be shortly described as follows. The CuO$_2$ plane electronic background
includes a defect subsystem bearing the doped holes besides the itinerant
hole poor material. Doping creates new defect states near the top of the
valence band ($\gamma$) in the charge-transfer gap. These states are
described by two subband components ($\alpha ,\beta$) representing the
"hot'' ($\pi ,0$)-type and "cold'' ($\frac{\pi}{2},\frac{\pi}{2}$)-type
regions of the momentum space. Bare normal state gaps between these
subbands and the valence band are supposed to be closed by progressive
doping (subband bottoms evolve down). The doping concentrations at which the
band overlaps are reached correspond to special (critical) points on the
phase diagram.

There are four regions of the doping in the model with specific dispositions
of the band components and the chemical potential: a) at very underdoping
the "young'' defect subsystem is gapped; b) in the underdoped region the
$\beta -\gamma$-bands overlap and $\mu$ is shifted down to intersect both
of them; c) the effectively doped region is headed by the optimal overlapping
of $\alpha -\beta -\gamma$ bands, all being intersected by the chemical
potential; d) at extended overdoping $\mu$ intersects only the overlapping
$\alpha -\gamma$ bands. The difference between the itinerant and defect
carriers becomes progressively washed up with doping. The $\alpha$ and
$\beta$ band densities of states are constant, for the valence band it
reduces with doping.

A plausible parameter set [23] is used for illustrative calculations. The
doped hole concentrations are scaled to $T_c(max)$ being reached at
$p=0.16$.

The effective Hamiltonian with the interband pairing channel between the
defect and itinerant states reads
\begin{equation}
H=\sum_{\sigma ,\vec{k},s}\epsilon_{\sigma}(\vec{k})
a\sp +_{\sigma ,\vec{k},s}a_{\sigma ,\vec{k},s}+
W\sum_{\sigma ,\sigma '}{}'\sum_{\vec{k},\vec{k}'}\sum_{\vec{q}}
a\sp +_{\sigma\vec{k}\uparrow}a\sp +_{\sigma (-\vec{k}+\vec{q})\downarrow}
a_{\sigma '(-\vec{k}'+\vec{q})\downarrow}
a_{\sigma '\vec{k}'\uparrow}\; .
\end{equation}

Here $\epsilon_{\alpha}=\xi_{\sigma}-\mu$, $s$ is the spin index,
$\sigma$ counts the bands and $\vec{q}$ is the pair momentum with the
components from the same bands. The superconductivity gap system
corresponding to (8) is
\begin{eqnarray}
\Delta_{\gamma}=W\sum_{\vec{k},\tau}{}\sp{\tau}\Delta_{\tau}(\vec{k})
E_{\tau}\sp{-1}(\vec{k}) th\frac{E_{\tau}(\vec{k})}{2\Theta}\\
\nonumber \Delta_{\tau}=W\sum_{\vec{k}}\Delta_{\gamma}(\vec{k})
E_{\gamma}\sp{-1}(\vec{k}) th\frac{E_{\gamma}(\vec{k})}{2\Theta}
\end{eqnarray}
with the usual form of the quasiparticle energies
$E_{\sigma}(\vec{k})=\pm\sqrt{\epsilon_{\sigma}\sp 2(\vec{k})
+\Delta_{\sigma}\sp 2(\vec{k})}$ and $\theta =k_BT$.

Here $\sum {}\sp{\tau}$ means the integration  over the different
energy intervals corresponding to the defect
system subbands $\tau =\alpha$, $\beta$.
The density of the paired carriers is
\begin{equation}
n_{s}=\frac{1}{2}\left[ \sum_{\vec{k}}\frac{\Delta\sp 2_{\sigma}(\vec{k})}
{E\sp 2_{\sigma}(\vec{k})} th\sp 2\frac{E_{\gamma}(\vec{k})}{2\Theta}+
\sum_{\vec{k}}{}\sp{\tau}\frac{\Delta\sp 2_{\tau}(\vec{k})}
{E\sp 2_{\tau}(\vec{k})} th\sp 2\frac{E_{\tau}(\vec{k})}{2\Theta}
\right]\; .
\end{equation}

The free energy corresponding to the (1) has been calculated analogously
to [18]. Our two-component system possesses two order parameters [18]
of different criticality. The "soft'' one of the Goldstone in phase type
is connected to amplitudes of the two superfluids. It determines both the
superconducting gaps and behaves critically at $T_c$ with the corresponding
critical coherence length characterizing the fluctuations. The
paired carrier effective mass associated with this mode reads
\begin{equation}
m_{ab}=\frac{1}{2} \frac{(\eta_{\alpha}+\eta_{\beta}+\eta_{\gamma})
(\delta_{\alpha}+\delta_{\beta}+\delta_{\gamma})}
{(\eta_{\alpha}+\eta_{\beta})\delta_{\gamma}m_{\gamma}\sp{-1}+
\eta_{\gamma}(\delta_{\alpha}m_{\alpha}\sp{-1}+\delta_{\beta}m_{\beta}\sp{-1})}\; ,
\end{equation}
Here the band effective masses are determined by the corresponding densities
of states. For $\mu -s$ being not too close to limiting
energies $\Gamma_{0\sigma}$ and $\Gamma_{c\sigma}$ of the bands one has
\begin{equation}
\eta_{\sigma}=W\rho_{\sigma}\ln \left[\left( \frac{2\gamma}{\pi}\right)\sp 2
\theta_c\sp{-2}| \Gamma_{0\sigma}-\mu||\Gamma_{c\sigma}-\mu|\right] \; ,
\end{equation}
\begin{equation}
\delta_{\sigma}=\frac{7}{2}\zeta (3)W\rho_{\sigma}|\mu -\Gamma_{0\sigma}|
(\pi\theta_c)\sp{-2}\; ,
\end{equation}
when $\mu$ is located in the integration region ($\zeta (x)$ is the
zeta-function; $\gamma =\exp (0.577)$).

If $\mu$ lies out of the band $\delta_{\sigma}=0$ and
\begin{equation}
\eta_{\sigma}=W\rho_{\sigma}\ln \left|
\frac{\Gamma_{c\sigma}-\mu}{\Gamma_{0\sigma}-\mu}\right|\; .
\end{equation}

\section{Isotope effects dependences on doping}
The transition temperature and supercarrier densitiy isotope effect exponents
have been calculated numerically using Expr.(9) and (10). $W$ has been
varied according to (6) and the necessary derivatives have been determined
from the doping dependence curves.

The calculated $\alpha (p)$ curve corresponding to $Z=0.05$ is shown
in Fig.1. The decrease of $\alpha$ with increasing $T_c$ is
illustrated. The major result consists in that even a contribution
of some percent from the electron-phonon interaction to the
interband pairing channel can lead to "normal valued'' transition
temperature isotope exponents. This can be understood from the
approximate formulae for $T_c$ of the present approach [16]. Here an
electron scale energy is cut off by an exponential factor containing
$V\sim M\sp{-1}$. Then $\alpha$ is found to be large for stronger
cut-off.

The experimental doping dependence of $\alpha$ in cuprates has been
approximated in [28] by the expression $\alpha =0.25\;
T_cT_{cm}\sp{-1} \sqrt{1-T_cT_{cm}\sp{-1}}$ ($T_{cm}$ is the maximum
$T_c$) for the underdoped region. Our result is compared with this
curve in Fig.2 and is exposed as being to slow. One source for this
discrepancy can be the dependence of $Z$ on doping. The enhancement
of the electron-phonon relative contribution into $W$ with
progressive underdoping can improve the result.

The isotope effect behaviour for the supercarrier density $n_s$ is
of the same nature as $\alpha (p)$ because the $T_c$ bell-like dependence
is driven by the supercarrier density. It changes according to the
efficiency of the interband pairing with varying $\mu$ and bands
disposition. The supercarrier fraction $n_sp$ behaviour explains in a
natural way the $T_c$ quenching at overdoping on the background of enhanced
doped hole concentration $p$ [29].

The paired carrier effective mass isotope effect exponent dependence
on doping is shown in Fig.3. The negative values of $\alpha_m$ at
underdoping agree with the observed trend find from superfluid
stiffness data supposing the absence of the contribution from the
paired carrier density [6,7]. Theoretical absolute values of
$\alpha_m$ are small as compared with $\alpha_n$ which is in the
same order as $\alpha$. As the result the overall negative isotope
exponent of the plane ($T=0$) penetration depth is determined by
both $\alpha_n$ and $\alpha_m$, see (4). The $\alpha_{\lambda}(p)$
theoretical curve is given in Fig.4. Its behaviour and observed
order of magnitude are determined by the isotope effect in the
paired carrier density overwhelming the paired carrier effective
mass contribution. This is in contrast with the widely accepted
assumption that the penetration depth is determined by the whole
amount of carriers present, which does not change at isotope
substitution [13,28], however, cf. [30]. The presence of paired
carrier density isotope effect is characteristic for multiband
superconductivity with interband pairing. There is an essential
difference with the conventional one-band case where all the (normal
state) carriers in the active momentum region will be paired at
$T=0$.

Experimentally the superfluid density oxygen isotope effect exponent
is estimated to be around 0.5. A representative value of
$\alpha_{\lambda}$ from Fig.4 leaves one for
$\alpha_\rho=-2\alpha_{\lambda}$ in this scale.

The present work shows that the electron-phonon interaction can cause in
multiband superconductors essential lattice-connected effects without
playing a leading role in the pairing mechanism. The paired carrier density
isotope effect must be taken into account at this.

This work was supported by the Estonian Science Foundation Grant No 6540.

\section*{References}
\begin{enumerate}
\item J.P.Franck in: Physical Properties of High-Temperature
Superconductors, ed. D.M.Ginzburg, World Sci., 1994, {\bf 4}, p. 189.\\
\item D.Zech et al., Nature {\bf 371}, 681 (1994).\\
\item G.Zhao et al., J. Phys. Cond. Mat. {\bf 10}, 9055 (1998). \\
\item G.Zaho, V.Kirtikar, D.E.Morris, Phys. Rev. B {\bf 63}, 220506(R)
(2001).\\
\item R.Khasanov, J. Phys. Cond. Mat. {\bf 16}, S 4439 (2004).\\
\item R.Khasanov et al., Phys. Rev. Lett. {\bf 92}, 057602 (2004). \\
\item J.L.Tallon et al., Phys. Rev. Lett. {\bf 94}, 237002 (2005). \\
\item A.Bianconi, J. Supercond. {\bf 18}, 625 (2005). \\
\item A.Bianconi, M.Filippi, in: Symmetry and Heterogeneity in High
Temperature Superconductors, ed. A.Bianconi, Springer, 2006, p. 21. \\
\item Phase Separation in Cuprate Superconductors, eds. K.A.M\"uller and
G.Benedek, World Sci., 1993.\\
\item K.A.M\"uller, Physica C {\bf 341-348}, 11 (2000). \\
\item S.J.L.Billinge, T.Egami, in: Lattice Effects in High-T$_c$
Superconductors, ed. Y.Bar-Yam, World Sci., 1992, p. 23. \\
\item T.Egami et al., Physica B {\bf 316-317}, 62 (2002).\\
\item H.Suhl, B.T.Matthias, L.R.Walker, Phys. Rev. Lett. {\bf 3}, 552 (1959).\\
\item V.A.Moskalenko, Fiz. Met. Metalloved. {\bf 8}, 503 (1959). \\
\item N.Kristoffel, P.Konsin, T.\"Ord, Riv. Nuovo Cimento {\bf 17}, 1 (1994).\\
\item P.Konsin, N.Kristoffel, T.\"Ord, Annalen der Physik {\bf 2}, 279 (1993). \\
\item T.\"Ord, N.Kristoffel, phys. stat. sol. b {\bf 216}, 1049 (1999). \\
\item A.S.Alexandrov, Physica C {\bf 363}, 231 (2001). \\
\item A.Bussmann-Holder et al., Europhys. Lett. {\bf 72}, 423 (2005). \\
\item T.Dahm, Phys. Rev. B {\bf 61}, 6381 (2000). \\
\item N.Kristoffel, P.Rubin, Solid State Commun. {\bf 122}, 265 (2002). \\
\item N.Kristoffel, P.Rubin, Physica C {\bf 402}, 257 (2004). \\
\item N.Kristoffel, P.Rubin, in: Symmetry and Heterogeneity in High
Temperature Superconductors, ed. A.Bianconi, Springer, 2006, p. 55. \\
\item N.Kristoffel, P.Rubin, Phys. Lett. A (2006). \\
\item N.Kristoffel, T.\"Ord, P.Rubin, Physica C (2006). \\
\item J.Kondo, Progr. Theor. Phys. {\bf 29}, 1 (1963). \\
\item T.Schneider, H.Keller, Phys. Rev. Lett. {\bf 69}, 3374 (1992). \\
\item N.Kristoffel, P.Rubin, J. Supercond. {\bf 18}, 705 (2005).\\
\item A.Bill, V.Z.Kresin, S.A.Wolf, Phys. Rev. B {\bf 57}, 10814 (1998).
\end{enumerate}

\newpage
\section*{Figure Captions}
$$
$$
Fig.1. The transition temperature isotope effect exponent vs doping. \\

Fig.2. The comparison of calculated $\alpha$ (full line) with the
averaged experimental data in the underdoped region. \\

Fig.3. The paired carrier effective mass isotope effect exponent on the
doping scale. \\

Fig.4. The penetration depth isotope effect exponent theoretical dependence
on doping.

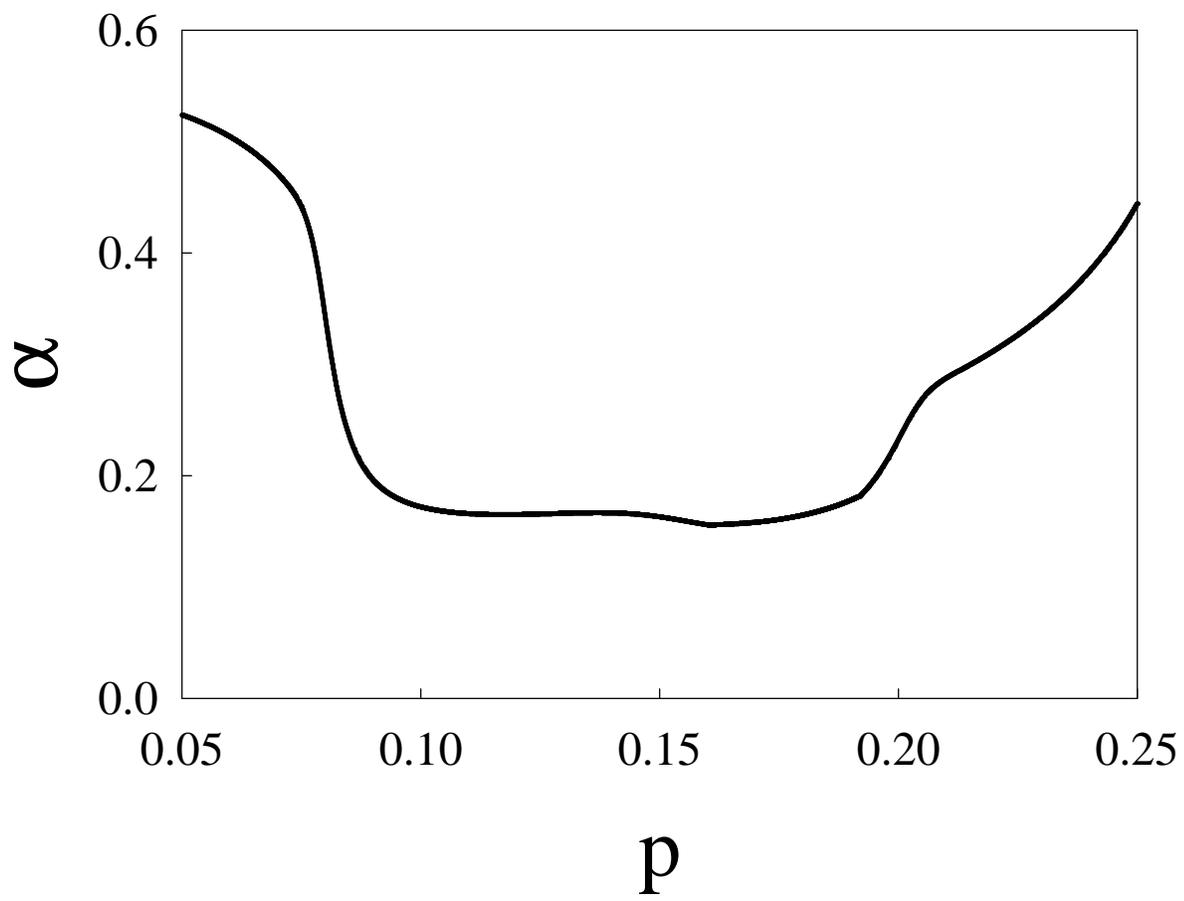
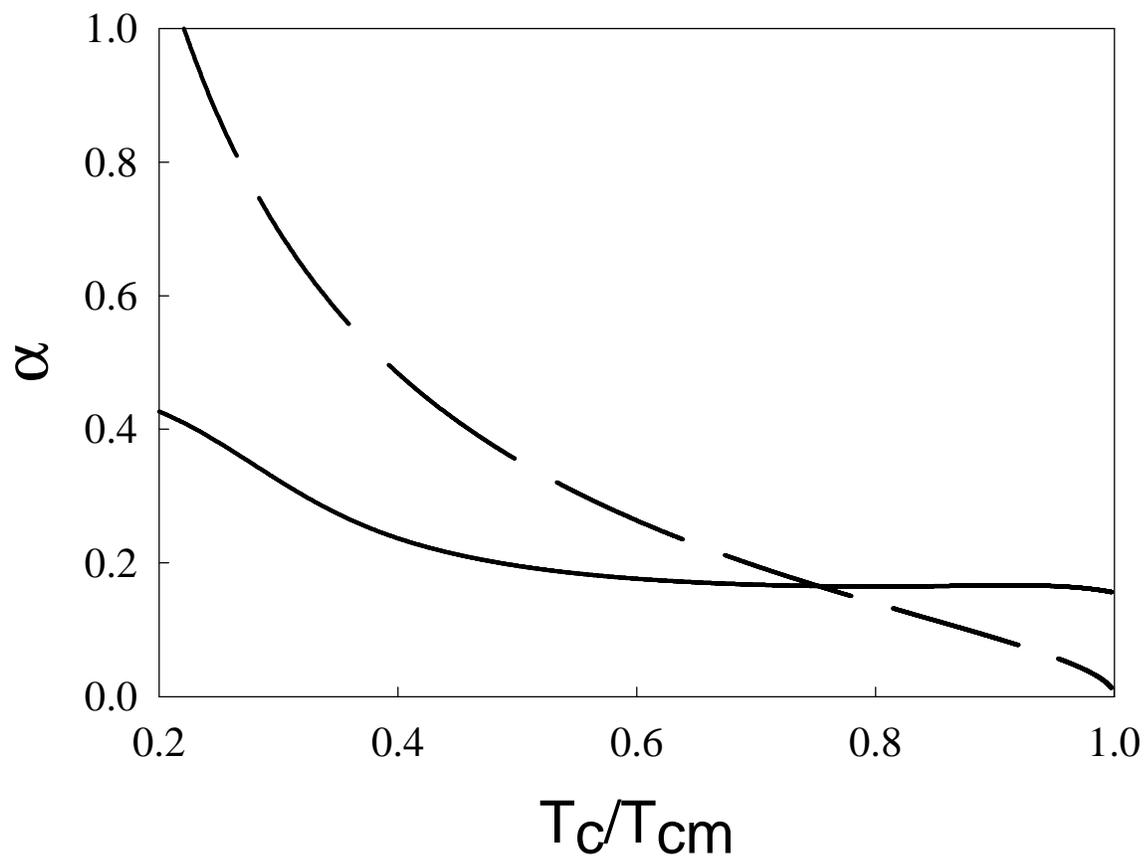
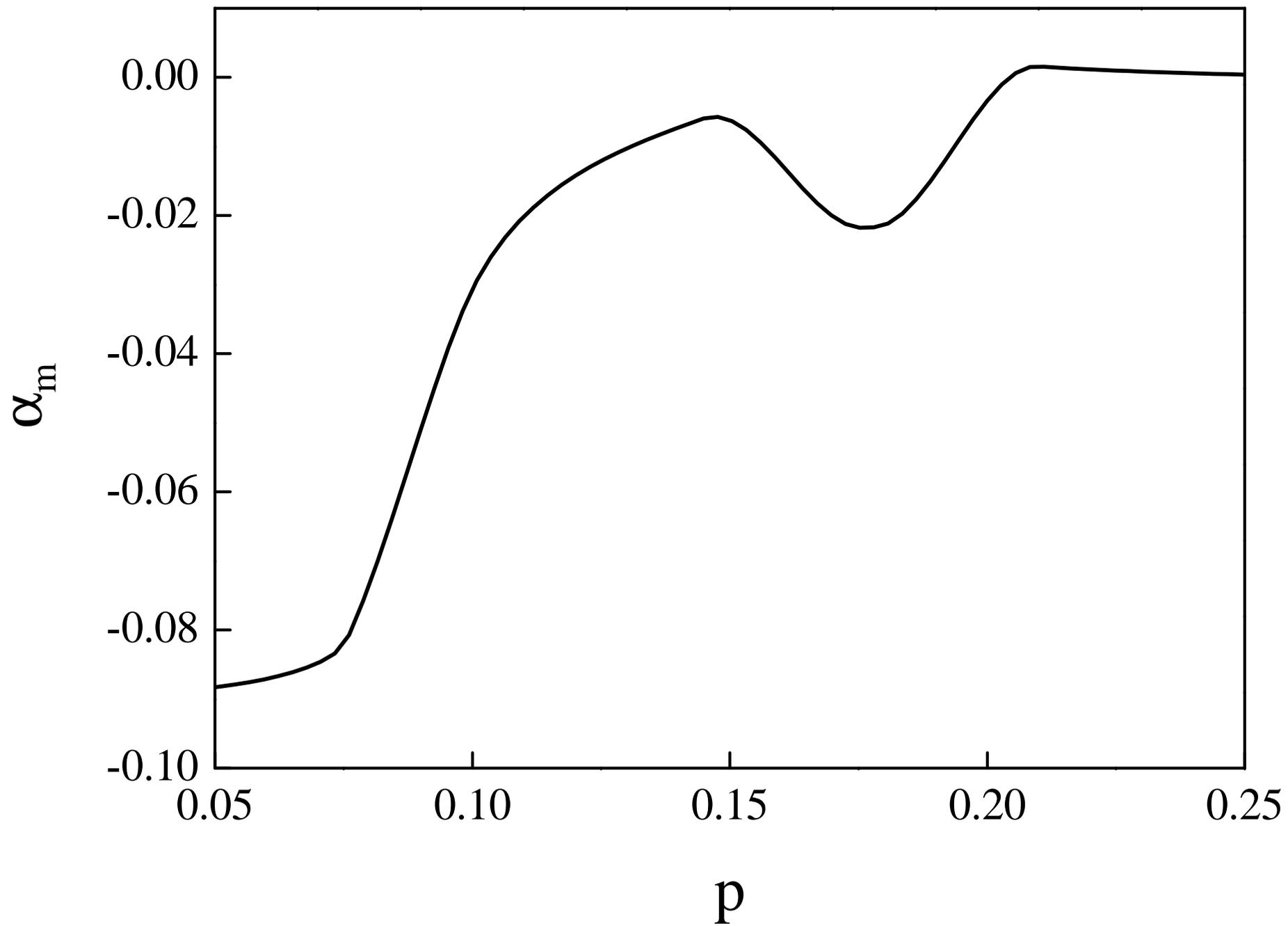
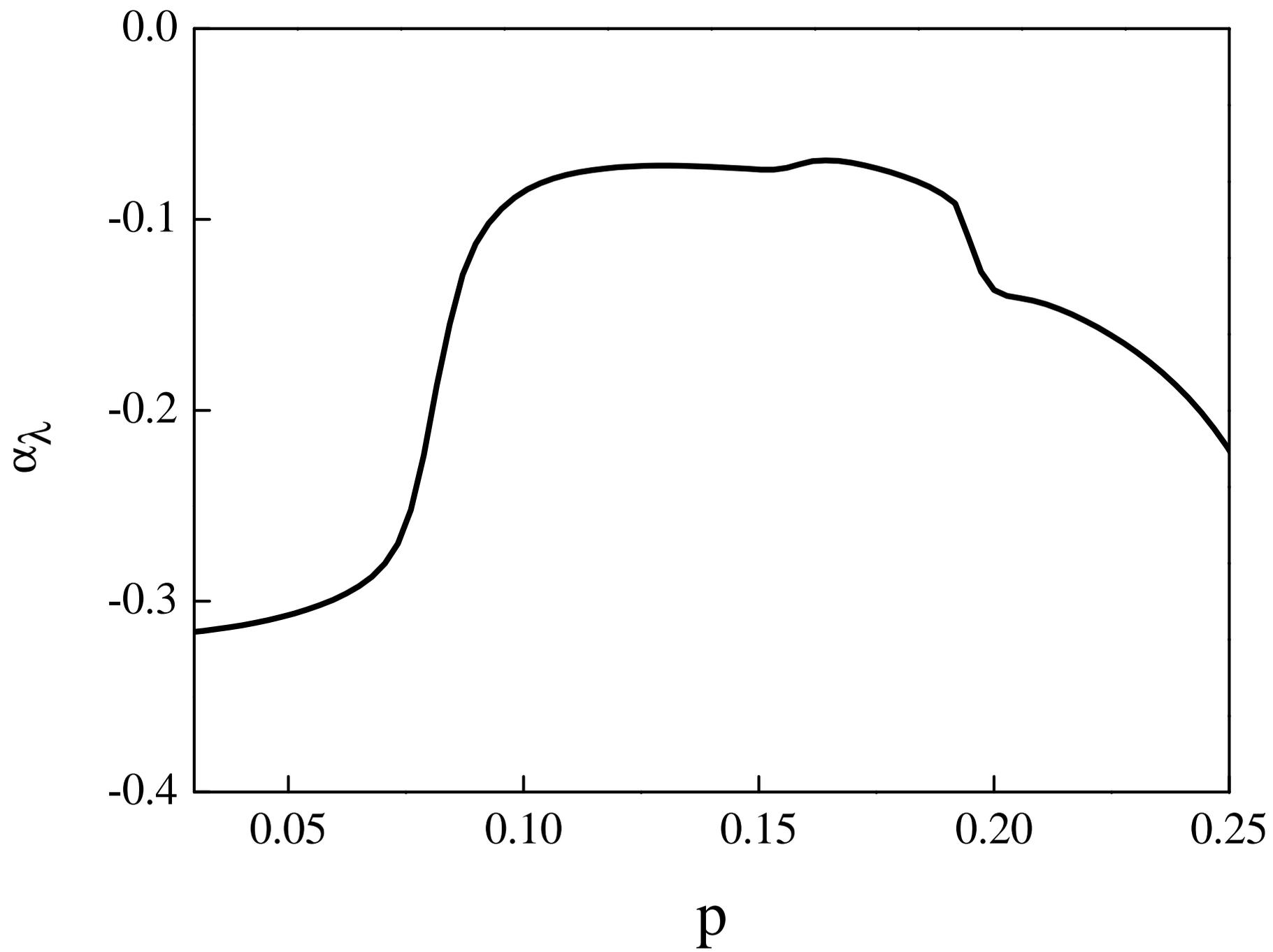
\end{document}